\documentclass[square]{article}

\usepackage[final]{d3s3_neurips_2024}

\usepackage[utf8]{inputenc} % allow utf-8 input
\usepackage[T1]{fontenc}    % use 8-bit T1 fonts
\usepackage{hyperref}       % hyperlinks
\usepackage{url}            % simple URL typesetting
\usepackage{booktabs}       % professional-quality tables
\usepackage{amsfonts}       % blackboard math symbols
\usepackage{nicefrac}       % compact symbols for 1/2, etc.
\usepackage{microtype}      % microtypography
\usepackage{xcolor}         % colors
\usepackage{graphicx}
\usepackage{amsthm,amsmath,amssymb}
\usepackage{mathrsfs}
\usepackage{subfigure}

\title{Generative Modeling and Data Augmentation for Power System Production Simulation}

\begin{document}
\author{%
  Linna Xu\\
  School of System Science and Engineering,Sun Yat-sen University\\
  Building 389, No. 135, Xingang West Road, Guangzhou \\
  \texttt{xuln6@mail2.sysu.edu.cn} \\
  \AND
  Yongli Zhu \\
  School of System Science and Engineering,Sun Yat-sen University\\
  Building 389, No. 135, Xingang West Road, Guangzhou \\
  \texttt{yzhu16@alum.utk.edu} \\
}

\maketitle

\begin{abstract}
As a key component of power system production simulation, load forecasting is critical for the stable operation of power systems. Machine learning methods prevail in this field. However, the limited training data can be a challenge. This paper proposes a generative model-assisted approach for load forecasting under small sample scenarios, consisting of two steps: expanding the dataset using a diffusion-based generative model and then training various machine learning regressors on the augmented dataset to identify the best performer. The expanded dataset significantly reduces forecasting errors compared to the original dataset, and the diffusion model outperforms the generative adversarial model by achieving about 200 times smaller errors and better alignment in latent data distributions.
\end{abstract}

\section{Introduction}

The modern power grid faces new challenges for stable and secure operation, such as the difficulty in forecasting load demand due to the uncertain charging profiles of electric vehicles (EV)'s. Accurate load forecasting informs power consumption for a given time horizon, enabling power utilities to schedule sufficient power generation while minimizing waste. Consequently, \textit{production simulation}, viz., an optimization program to economically allocate each power plant's output, is widely used by utility companies.

Various machine learning methods for load forecasting have been reported. Fan et al. \cite{ref1} employed a Long Short-Term Memory (LSTM) network for short-term load forecasting. Similarly, Kong et al. \cite{ref2} developed a hybrid model combining Convolutional Neural Networks (CNN) and LSTM to enhance forecasting accuracy by capturing spatial and temporal dependencies. A random forest (RF) model was utilized in \cite{ref3} to address overfitting through ensemble learning for short-term load forecasting. Wang et al. \cite{ref4} applied a Gradient Boosting Decision Tree (GBDT) model to effectively capture nonlinear relationships in the data. The CatBoost model was used in \cite{ref5} to predict power load demands, demonstrating strong performance with mixed data types.

The methods mentioned above assume abundant, high-quality load demand data, which is often unavailable in the power industry. Communication failures, device malfunctions, and newly built communities with limited data can impede accurate load forecasting.

Therefore, obtaining a large number of high-quality datasets from a given small dataset is a key issue here. Generative Adversarial Networks (GANs) have excelled in data generation\cite{ref7,ref8,ref9}, and many variants have been developed, among which TimeGAN (Time-series Generative Adversarial Networks) \cite{ref10} performs well in synthesizing time series data. In addition, the diffusion model \cite{ref11} has been applied for time-series data generation in recent years. For example, Yuan. et al. proposed an interpretable diffusion model for generic time series generation \cite{ref12}.

In this paper, we explore the effectiveness of load data augmentation for power system production simulation using TimeGAN and TS-Diffusion, respectively. The first step is augmenting the original dataset, the second step is training a forecasting model based on the augmented dataset, and the last step is feeding the predicted load demand in an optimization model to conduct the power system production simulation. The source code and data are freely accessible at \url{https://github.com/Becklishious/NeurIPS2024}.

\section{Dataset Augmentation for Load Forecasting}

In this section, we present a brief description of the original dataset and the TS-Diffusion model for data augmentation. The math details are described in Appendix A.2. The TimeGAN-based data augmentation model and the ExtraTree-based \cite{ref6} load forecasting model are described in Appendix A.1 and A.3.

\subsection{Dataset Description}

The dataset used in this paper is collected from a household in northwest China, comprising 168 hourly load records from April 16 to April 22, 2024. It includes additional meteorological data: temperature, barometric pressure, wind speed, wind direction, surface horizontal radiation, direct normal radiation, and diffuse radiation. These variables will be included during dataset augmentation. In our load forecasting model, the meteorological data serve as features, while the load values are used as labels.

\subsection{TS-Diffusion Model for Load Demand Augmentation}

\begin{figure}[htbp]
  \centering
  \includegraphics[width=1\linewidth]{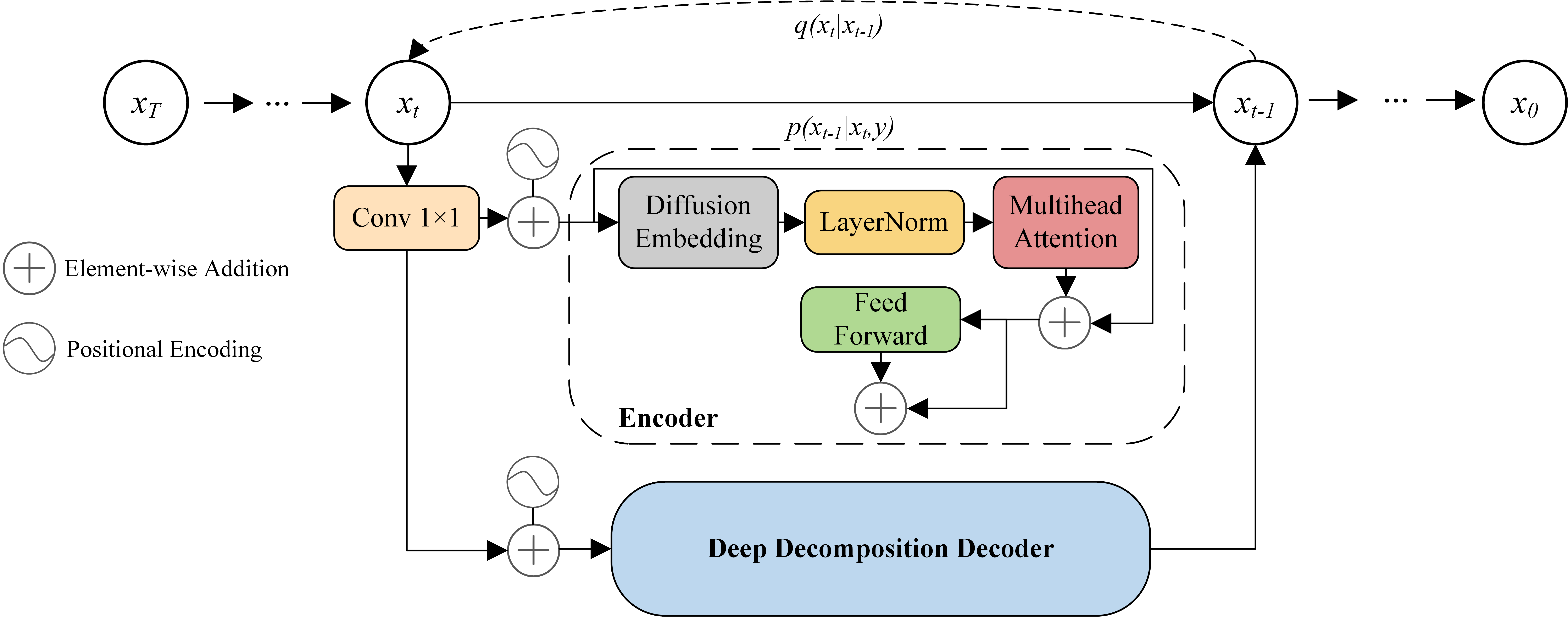}
  \caption{TS-Diffusion training framework.}
  \label{fig2}
\end{figure}

TS-Diffusion is a diffusion model for time series data generation based on a Transformer-inspired architecture combined with a decomposition design. TS-Diffusion performs well in tasks like missing-value interpolation. Hence, we adopt it to augment time series data. The training process diagram of the TS-Diffusion generation model is shown in Figure \ref{fig2}.

The encoder module processes the input time series using a multi-head attention mechanism and a feed-forward neural network. The decoder module also uses multi-head attention and feed-forward layers, plus a deep decomposition design to capture the trend and seasonality components of the time series. The diffusion-embedding module incorporates time-step information, and the positional encoding module adds positional information to help the model capture the inherent temporal patterns of the data.

\section{Experiment Results}

This section mainly 1) compares the \textit{quality} of the augmented datasets obtained by the two generative machine learning approaches and 2) gives a simple \textit{showcase} for power system production simulation. An in-depth analysis of the experiment results can be found in Appendix A.4. All the experiments are implemented on a desktop PC with Intel 5.4GHz CPU and 32GB RAM.

\subsection{Quality of TS-Diffusion Augmented Dataset}

 The original data is divided into the training and test sets at the beginning with an 8:2 split ratio. Then, the TS diffusion model is used to augment the original training set (about 134 data samples) to 3456 samples. Prediction models are then trained respectively on the original training set and the augmented training set. Besides, another comparison dataset is established by expanding the original training set to the same size as the TS diffusion-augmented dataset using simple replication, followed by model training. RMSE (Root Mean Square Error) and MAE (Mean Absolute Error) are used as the performance metrics. Four types of regression models are considered: ExtraTree, Random Forest, CatBoost, and XGBoost. Each model is independently trained once on the original, replicated, and augmented datasets and tested on the (previously split) 20\% testing set. Results are shown in Table~\ref{table1}.

\begin{table}[htbp]
  \caption{Performance comparison on the original, replicated and augmented datasets (TS-Diffusion)}
  \label{table1}
  \centering
  \begin{tabular}{llll}
    \toprule
    % \multicolumn{2}{c}{Part}                   \\
    % \cmidrule(r){1-2}
    Model        & dataset        & RMSE                   & MAE      \\
    \midrule
    XGBoost      & original       & 0.05774              & 0.04276   \\
                 & replicated     & 0.06485              & 0.04427   \\
                 & augmented      & 0.01526              & 0.00249   \\
    CatBoost     & original       & 0.04389              & 0.03323   \\
                 & replicated     & 0.04536              & 0.03243   \\
                 & augmented      & 0.00236              & 0.00098   \\          
    RandomForest & original       & 0.04183              & 0.02952   \\
                 & replicated     & 0.05846              & 0.03968   \\
                 & augmented      & 0.00153              & 0.00013   \\          
    ExtraTree    & original       & 0.04467              & 0.03209   \\
                 & replicated     & 0.04495              & 0.03229   \\
                 & augmented      & \textbf{0.00023}     & \textbf{0.00004}   \\                    
    \bottomrule
  \end{tabular}
\end{table}

There is no risk of data leakage because: 1) the original dataset has been divided into the training set and test set at the beginning 2) \textit{only} the training set is used to establish the GAN model and 3) the final augmented dataset contains no entry from the testing set. From Table~\ref{table1}, we can see that the results of the TS-Diffusion augmented dataset are better than the original and replicated datasets, and this conclusion holds for all four models. In particular, when the ExtraTree model is trained using the augmented dataset, it has the best prediction effect on the testing set, with an RMSE of 0.00023 and an MAE of 0.00004, which cannot be achieved by simply duplicating the original dataset.

\subsection{Quality of TimeGAN Augmented Dataset}  \label{sec32}

Similar to 3.1, we use TimeGAN to augment the dataset. The experiment results on the augmented dataset are shown in Table~\ref{table2} (comparisons with other datasets are put in Table~\ref{table3} of the Appendix).

\begin{table}[htbp]
  \caption{Performance on the augmented dataset (TimeGAN)}
  \label{table2}
  \centering
  \begin{tabular}{llll}
    \toprule
    % \multicolumn{2}{c}{Part}                   \\
    % \cmidrule(r){1-2}
    Model        & dataset        & RMSE                   & MAE         \\
    \midrule
    XGBoost      & augmented      & 0.06398              & 0.03445   \\
    CatBoost     & augmented      & 0.05637              & 0.02761   \\          
    RandomForest & augmented      & 0.06130              & 0.02949   \\          
    ExtraTree    & augmented      & \textbf{0.05356}     & \textbf{0.02395}   \\                    
    \bottomrule
  \end{tabular}
\end{table}

From Table~\ref{table2}, we can find that the ExtraTree model still performs best. However, the RMSE increases (compared to the prediction model trained on the original data). Therefore, the robustness of the TimeGAN-based data-augmentation model is slightly inferior to that of the TS-diffusion model.

\subsection{A Simple Showcase for Power System Production Simulation}

The previously predicted load data can be utilized in a standard power system production simulation procedure, which means solving the following optimization model:

Suppose a region's load is supplied by grid-purchased power $P_{grid}(t)$ and photovoltaic (PV) power $P_{pv}(t)$ . The cost of PV power is 0.4 \$/kWh, while the cost of the grid-purchased power is 1 \$/kWh. To optimize this region's power operation, the objective is to minimize the power production cost while meeting load requirements $P_{load}(t)$. The objective function is defined in Eq.(\ref{eq7}), with constraints from Eq.(\ref{eq8}) to Eq.(\ref{eq10}).

 \vspace{-5mm}
 
\begin{align}
  \label{eq7} &\text{minimize} & & cost_{grid}\sum_{t=1}^{T=24}P_{grid}(t)+cost_{pv}\sum_{t=1}^{T=24}P_{pv}(t)  \\
  \label{eq8} &\text{subject to} & & P_{load}(t)=P_{grid}(t)+P_{pv}(t), &t=1,...,T\\
  \label{eq9} & & & P_{grid}(t) \geq 0, &t=1,...,T\\
  \label{eq10} & & & 0 \leq P_{pv}(t) \leq P_{pv,max}(t), &t=1,...,T
\end{align}

\begin{figure}
	\centering
	\subfigure[]
        {
		\begin{minipage}[t]{0.31\textwidth}
			\includegraphics[width=0.9\textwidth]{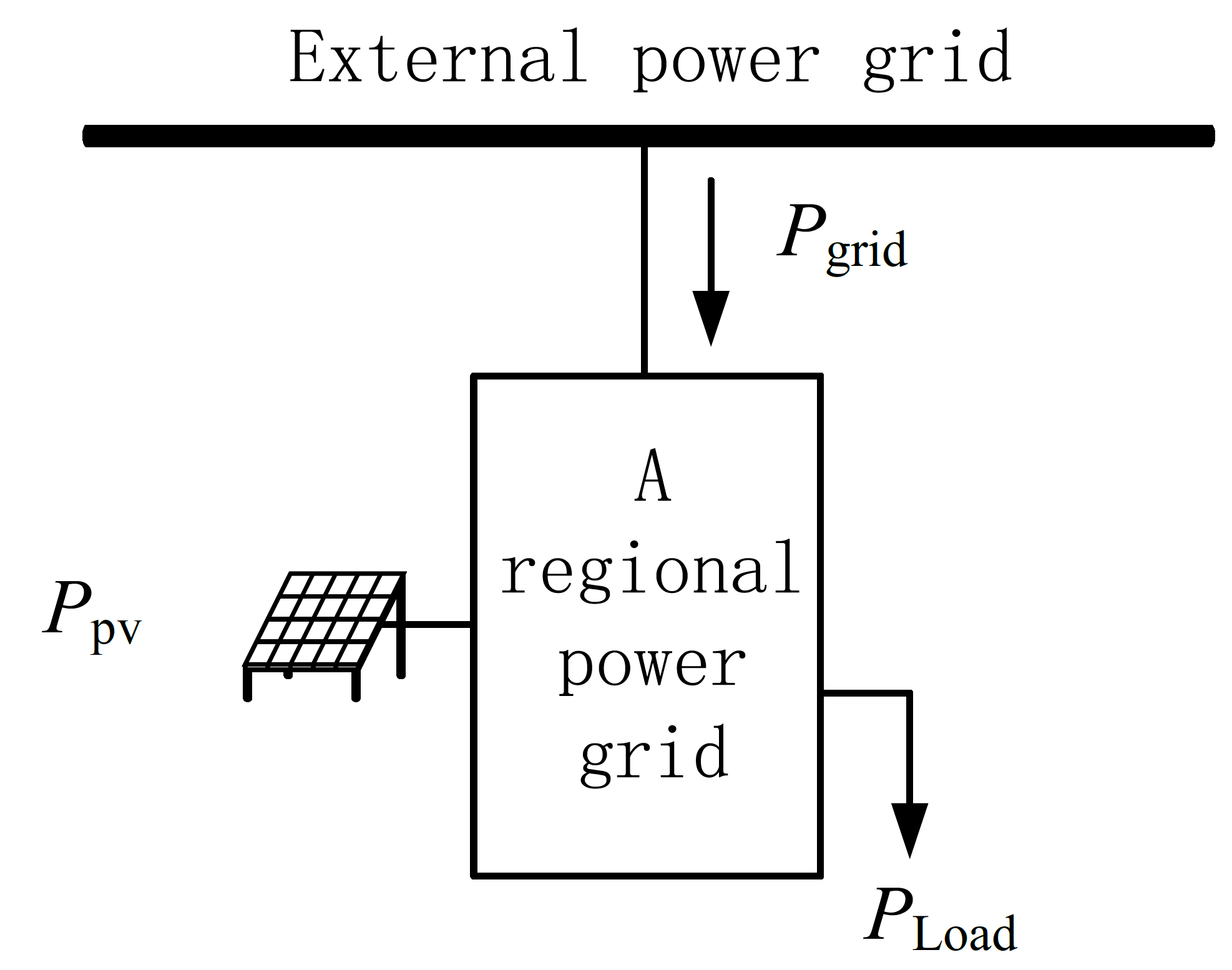}
		\end{minipage}
		\label{fig7.1}
	}
        \subfigure[]
        {
        \begin{minipage}[b]{0.65\textwidth}
   		 	\includegraphics[width=0.9\textwidth]{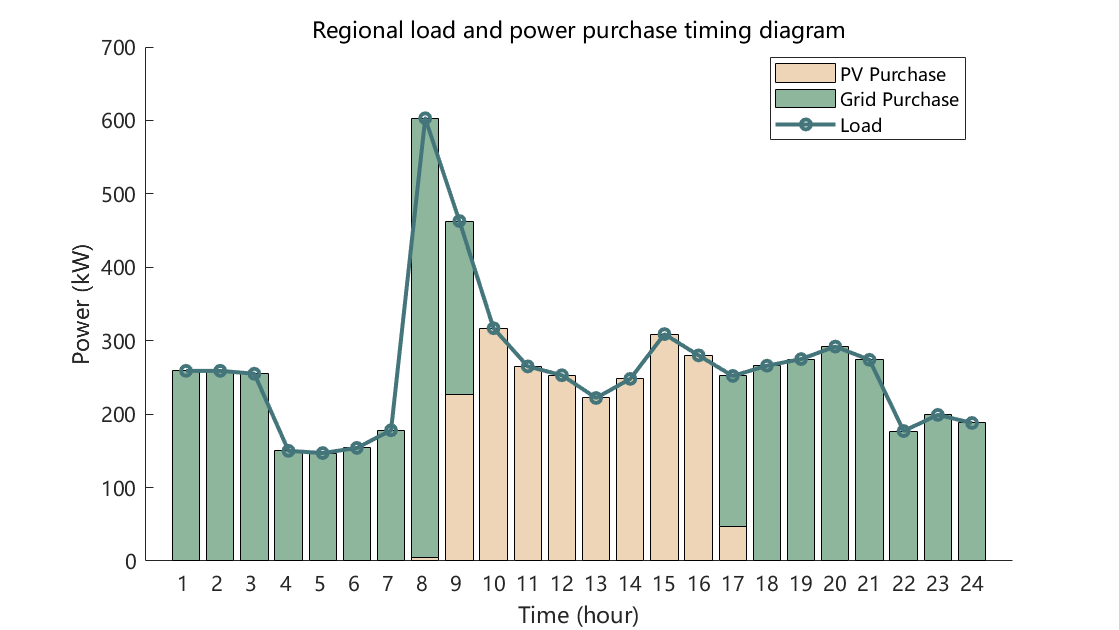}
        \end{minipage}
		\label{fig7.2}
        }
	\caption{(a) a regional power grid; ~ (b) production simulation results}
	\label{fig7}
\end{figure}

In Eq.(\ref{eq10}), $P_{pv,max}(t)$ is the maximum possible PV power at time $t$. Since the load demand of the future horizon (e.g., the next day) is unknown, we employ the ExtraTree-based load forecasting model (trained on the dataset augmented by TS-Diffusion) ``as'' the future day’s load, $P_{load}(t)$. Then, this forecast load will be substituted in Eq.(\ref{eq8}). After solving the optimization problem, the simulation results of the PV generation and grid power purchase are depicted in Figure \ref{fig7}.

% \begin{figure}[htbp]
%   \centering
%   \includegraphics[width=0.7\linewidth]{power_plan.png}
%   % \fbox{\rule[-.5cm]{0cm}{4cm} \rule[-.5cm]{4cm}{0cm}}
%   \caption{Simulation result of power system production}
%   \label{fig7}
% \end{figure}

During the no-PV-power period (before sunrise or after sunset), the region's power supply fully relies on external grid support. As the PV power increases, the grid power exchange gradually decreases. From 10:00 to 16:00, when the PV power is the most sufficient, it fully meets the region's load needs.

\section{Conclusion and Future Work}

In this paper, we propose a method to improve the accuracy of load forecasting models using generative machine learning under small samples. The quality of the generated load data ( especially by the diffusion model) significantly improves the load-forecasting accuracy, demonstrating the feasibility and capability of generative machine learning for power system production simulation. 

Future work includes fine-tuning the generative models for better data quality and conducting additional comparisons. A limitation of this study is that the trained generative model for one regional power system may not directly apply to neighboring regions. Thus, applying transfer learning to enhance the model's generalizability will be the next step.

\newpage
\appendix

\section{Appendix}

\subsection{GAN-based Load Demand Augmentation Model}

TimeGAN (Time-series Generative Adversarial Networks) is a GAN model customized for time series data generation. It integrates GAN with self-supervised learning to capture complex temporal patterns. Like traditional GAN, it includes a generator that produces synthetic data and a discriminator that distinguishes between real and generated data. Through iterative training, the generator improves in producing simulated time series.

\begin{figure}[htbp]
  \centering
  \includegraphics[width=1\linewidth]{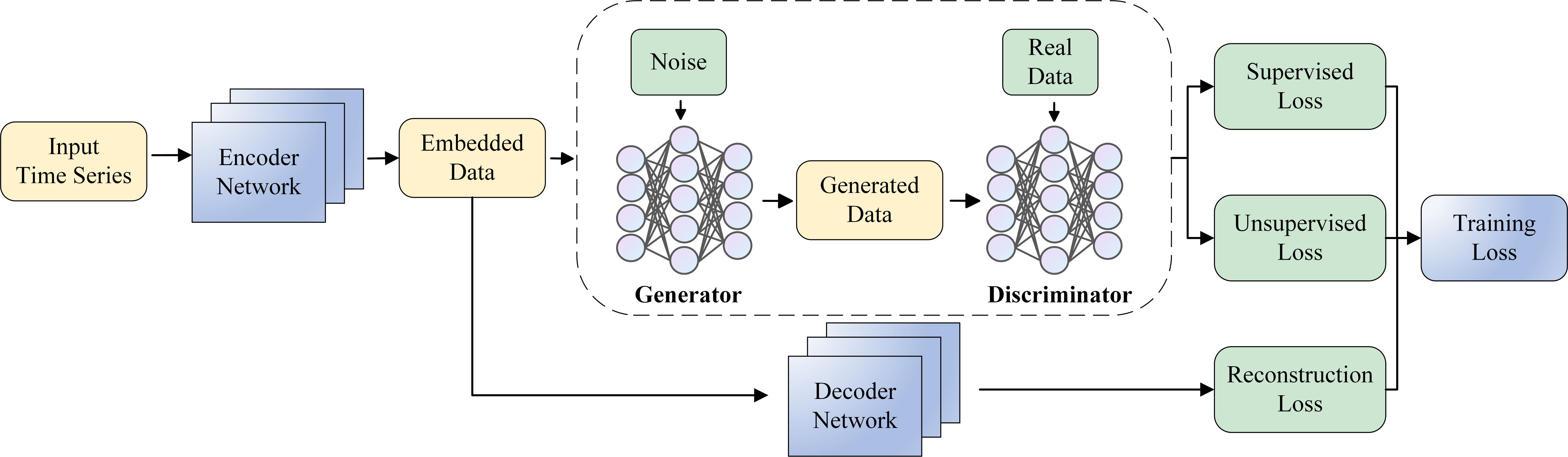}
  % \fbox{\rule[-.5cm]{0cm}{4cm} \rule[-.5cm]{4cm}{0cm}}
  \caption{TimeGAN training framework.}
  \label{fig1}
\end{figure}

A key feature of TimeGAN is its use of self-supervised learning via an auto-encoder, comprising an encoder that maps time series data to latent representations and a decoder that reconstructs the original data. This structure enhances the model's ability to capture the intrinsic time-series features. Figure \ref{fig1} illustrates the TimeGAN training process.

In Figure \ref{fig1}, the original data $\mathbf{s},\mathbf{x}_{1:T}$ is processed by the encoder and decoder networks to obtain the latent representation $\mathbf{h}_{\mathcal{S}},\mathbf{h}_{1:T}$ and the reconstruction values $\tilde{\mathbf{s}},\tilde{\mathbf{x}}_{1:T}$. The reconstruction loss is defined by Eq. (\ref{eq1}).

\begin{equation}\label{eq1} 
  \mathcal{L}_{\mathrm{R}}=\mathbb{E}_{\mathbf{s},\mathbf{x}_{1:T}\sim p}\big[\|\mathbf{s}-\tilde{\mathbf{s}}\|_{2}+\sum_{t}\|\mathbf{x}_{t}-\tilde{\mathbf{x}}_{t}\|_{2}\big]
\end{equation}

When TimeGAN generates data, the generator receives two types of inputs during training. First, in the open-loop mode, the generator receives synthetic embedding $\hat{\mathbf{h}}_{\mathcal{S}},\hat{\mathbf{h}}_{1:t-1}$ to generate the next synthetic vector $\hat{\mathbf{h}}_t$. The gradient is calculated based on the unsupervised loss in Eq.(\ref{eq2}), in order to provide the correct classification results $\hat{y}_\mathcal{S},\hat{y}_{1:T}$ for the generated data and the training data as much as possible.

\begin{equation}\label{eq2} 
  \mathcal{L}_{\mathrm{U}}=\mathbb{E}_{\mathbf{s},\mathbf{x}_{1:T}\sim p}\Big[\log y_{\mathcal{S}}+\sum_{t}\log y_{t}\Big]+\mathbb{E}_{\mathbf{s},\mathbf{x}_{1:T}\sim\hat{p}}\Big[\log(1-\hat{y}_{\mathcal{S}})+\sum_{t}\log(1-\hat{y}_{t})\Big]
\end{equation}

In the closed-loop mode, the generator receives sequences of actual-data embeddings $\mathbf{h}_{1:t-1}$ to generate the next latent vector. The gradient is then calculated based on the supervised loss of Eq.(\ref{eq3}) to capture the difference between the distributions of the real data and generated data.

\begin{equation}\label{eq3} 
  \mathcal{L}_{\mathrm{S}}=\mathbb{E}_{\mathbf{s},\mathbf{x}_{1:T}\sim p}\big[\sum_{t}\|\mathbf{h}_{t}-g_{\mathcal{X}}(\mathbf{h}_{\mathcal{S}},\mathbf{h}_{t-1},\mathbf{z}_{t})\|_{2}\big]
\end{equation}

Compared with traditional GANs, TimeGAN can capture temporal dependencies and generate sequences with a similar temporal dependency structure as real data. In addition, by combining adversarial training and self-supervised learning, TimeGAN can simultaneously optimize the processes of data generation and feature transformation. Hence, high-quality data can be generated. Moreover, the trained TimeGAN can generate multi-dimensional time series of arbitrary lengths.

\subsection{TS-Diffusion-based Load Demand Augmentation Model}

As shown in Figure \ref{fig2}, the TS-Diffusion model contains forward and reserve processes. In this setting, a sample from the data distribution $x_0 \sim q(x)$ is gradually noised by a Gaussian noise $\mathcal{N}$ during the forward process, where the transition is parameterized by $q(x_t \textbar x_{t-1}) = \mathcal{N} (x_t;\sqrt{1-\beta_t} x_{t-1},\beta_t \mathbf{I})$ with $\beta_t \in (0,1)$ as the amount of noise added at diffusion step $t$. Then a neural network learns the reverse process of gradual denoising the sample via reverse transition $p_{\theta} (x_{t-1} \textbar x_t) = \mathcal{N} (x_{t-1}; \mu_{\theta}(x_t,t),\sum_{\theta}(x_t,t))$. The reverse process can be approximated via Eq. (\ref{eq4}).

\begin{equation}\label{eq4} 
  x_{t-1}=\frac{\sqrt{\bar{\alpha}_{t-1}}\beta_{t}}{1-\bar{\alpha}_{t}}\hat{x}_{0}(x_{t},t,\theta)+\frac{\sqrt{\alpha_{t}}(1-\bar{\alpha}_{t-1})}{1-\bar{\alpha}_{t}}x_{t}+\frac{1-\bar{\alpha}_{t-1}}{1-\bar{\alpha}_{t}}\beta_{t}z_{t}
\end{equation}

where $z_{t}\sim\mathcal{N}(0,\mathbf{I}),\alpha_{t}=1-\beta_{t}$ and $\bar{\alpha}_{t}=\prod_{s=1}^{t}\alpha_{s}$.TS-Diffusion trained this denoising model $\mu_{\theta}(x_t,t)$ using a weighted mean squared error loss, the reweighting strategy is shown in Eq. (\ref{eq5}).

\begin{equation}\label{eq5} 
  \mathcal{L}_{simple}=\mathbb{E}_{t,x_0}\left[w_t\|x_0-\hat{x}_0(x_t,t,\theta)\|^2\right], w_t=\frac{\lambda\alpha_t(1-\bar{\alpha}_t)}{\beta_t^2}
\end{equation}

where $\lambda$ is a constant. These loss terms are down-weighted at small $t$ to force the network focus on a larger diffusion step. In addition, TS-Diffusion guides an interpretable diffusion training by applying the Fourier transformation in the frequency domain, i.e.,

\begin{equation}\label{eq6} 
  \mathcal{L}_{\theta}=\mathbb{E}_{t,x_{0}}\left[w_{t}\left[\lambda_{1}\|x_{0}-\hat{x}_{0}(x_{t},t,\theta)\|^{2}+\lambda_{2}\|\mathcal{FFT}(x_{0})-\mathcal{FFT}(\hat{x}_{0}(x_{t},t,\theta))\|^{2}\right]\right]
\end{equation}

where $\mathcal{FFT}$ denotes the Fast Fourier Transformation, and $\lambda_1,\lambda_2$ are the balancing weights for the two losses in Eq. (\ref{eq6}).

\subsection{ExtraTree-based Load Forecasting Model}

ExtraTree is a decision tree-based machine learning model that improves its generalization ability and computational efficiency by introducing \textit{ extreme stochasticity} to randomly select features and feature splitting points.

\begin{figure}[htbp]
  \centering
  \includegraphics[width=1\linewidth]{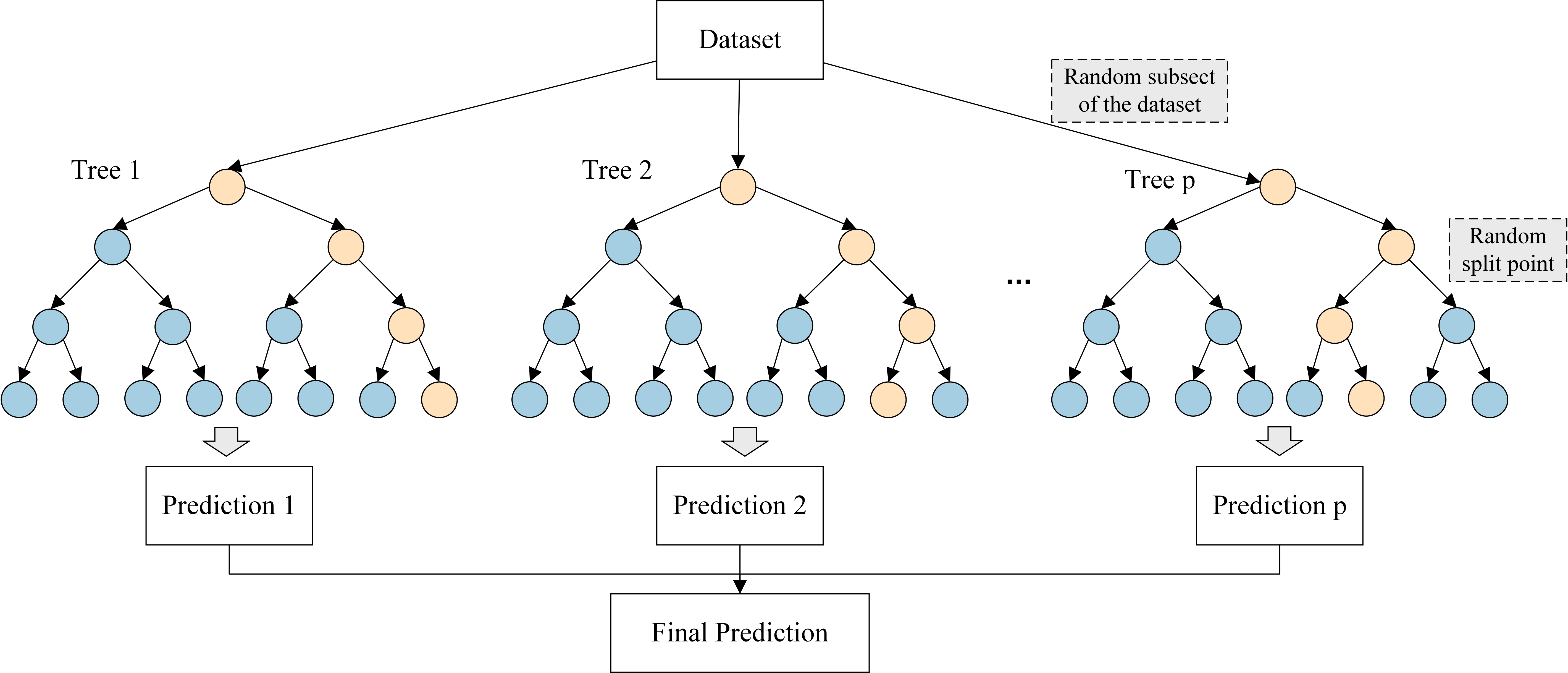}
  % \fbox{\rule[-.5cm]{0cm}{4cm} \rule[-.5cm]{4cm}{0cm}}
  \caption{ExtraTree Model.}
  \label{fig3}
\end{figure}

Unlike the usual approach that uses only a subset of the data, it uses the entire training dataset to build each tree. During the training process, the ExtraTree model introduces a lot of randomness. At each split node, it randomly selects features and feature values to split rather than choosing the optimal feature split point. More specifically, it randomly selects a subset of all features, then randomly selects a feature from this subset to split, and randomly selects one of the possible split points as the actual split point. Figure \ref{fig3} illustrates the basic idea of the ExtraTree model for load forecasting.

When making predictions, the ExtraTree model first makes individual predictions for each tree. Specifically, the input data is passed to each decision tree, and each tree makes a prediction based on the feature splitting rule for the path from the root node to the leaf nodes. Finally, the predictions of all decision trees are averaged to obtain the final prediction.

\subsection{An In-depth Analysis of the Experiment Results}
In this subsection, the "generation quality" of the TS-Diffusion and TimeGAN augmented datasets are respectively analyzed. The load forecasting models are trained on the augmented datasets, of which the MAE and RMSE are compared with the forecasting models trained on the original data. The complete results of section \ref{sec32} are shown in Table \ref{table3}.

\begin{table}[htbp]
  \caption{Performance comparison on the original, replicated and augmented datasets (TimeGAN)}
  \label{table3}
  \centering
  \begin{tabular}{llll}
    \toprule
    % \multicolumn{2}{c}{Part}                   \\
    % \cmidrule(r){1-2}
    Model        & dataset        & RMSE                   & MAE      \\
    \midrule
    XGBoost      & original       & 0.05774              & 0.04276   \\
                 & replicated     & 0.06485              & 0.04427   \\
                 & augmented      & 0.06398              & 0.03445   \\
    CatBoost     & original       & 0.04389              & 0.03323   \\
                 & replicated     & 0.04536              & 0.03243   \\
                 & augmented      & 0.05637              & 0.02761   \\          
    RandomForest & original       & 0.04183              & 0.02952   \\
                 & replicated     & 0.05846              & 0.03968   \\
                 & augmented      & 0.06130              & 0.02949   \\          
    ExtraTree    & original       & 0.04467              & 0.03209   \\
                 & replicated     & 0.04495              & 0.03229   \\
                 & augmented      & \textbf{0.05356}     & \textbf{0.02395}   \\                    
    \bottomrule
  \end{tabular}
\end{table}

The results in Table~\ref{table3} reveal that the model trained on the TimeGAN-augmented dataset outperforms the one trained on the original and replicated data regarding the MAE but underperformed in RMSE. MAE measures the average absolute error between predicted and actual values, giving equal weight to each error. In contrast, the RMSE is more sensitive to larger errors due to the squaring operation. The augmented data leads to lower MAE but higher RMSE, indicating smaller errors overall with a few extreme outliers. This suggests that TimeGAN may have introduced slight anomalies during augmentation, thus enlarging the RMSE. Besides, the augmented data possibly deviates from the original data distribution in the tail, leading to poorer model performance in extreme cases. While the augmentation improves prediction accuracy to some extent, it somewhat compromises robustness.

On the other hand, the load forecasting models trained on the TS-Diffusion augmented dataset are better than those trained on the original data, both in terms of MAE and RMSE. Also, compared with the TimeGAN augmented dataset, the quality of the TS-Diffusion augmented dataset is obviously better. Taking ExtraTree as an example, the MAEs of the model trained on the original dataset, the TimeGAN augmented dataset, and the TS-Diffusion augmented dataset are respectively 0.03209, 0.02395, and 0.00004. The performance of the TS-Diffusion augmented dataset is remarkable.

\begin{figure}[htbp]
  \centering
  \includegraphics[width=0.9\linewidth]{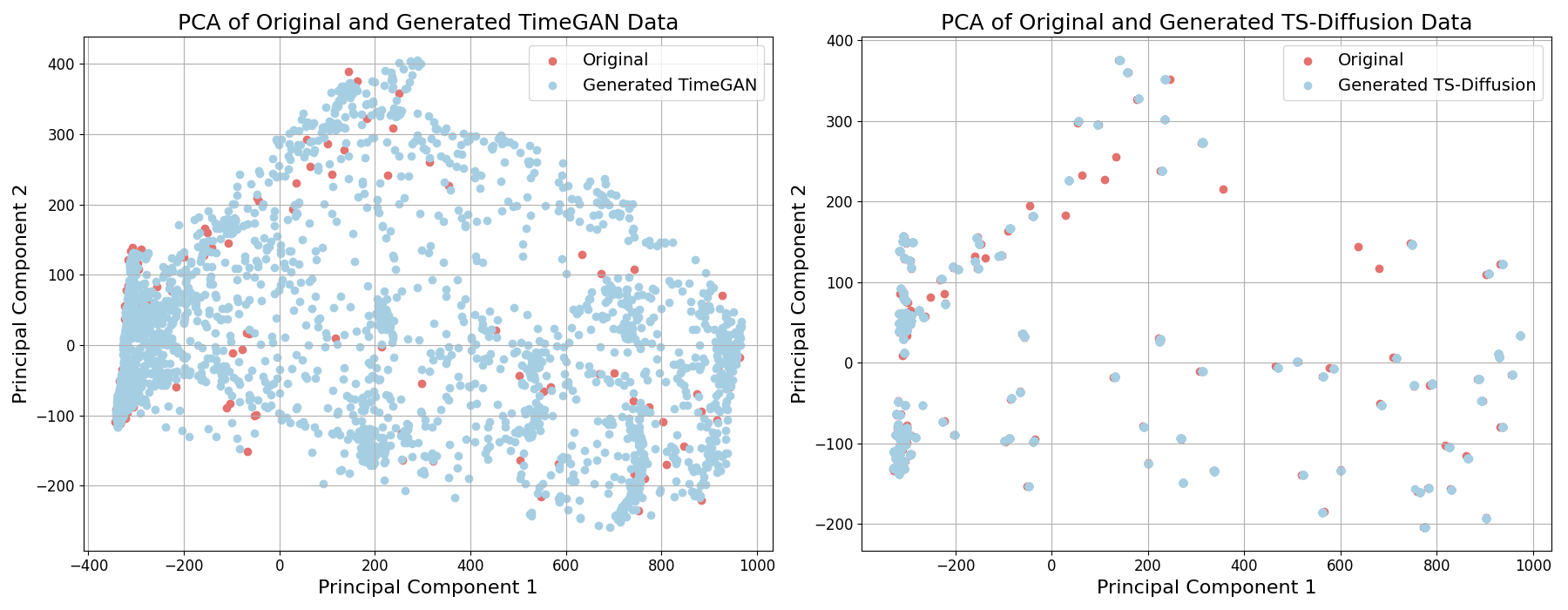}
  \caption{PCA of the TS-Diffusion generated data and the original data (right) and PCA of the TimeGAN generated data and the original data (left).}
  \label{fig4}
\end{figure}

\begin{figure}[htbp]
  \centering
  \includegraphics[width=0.9\linewidth]{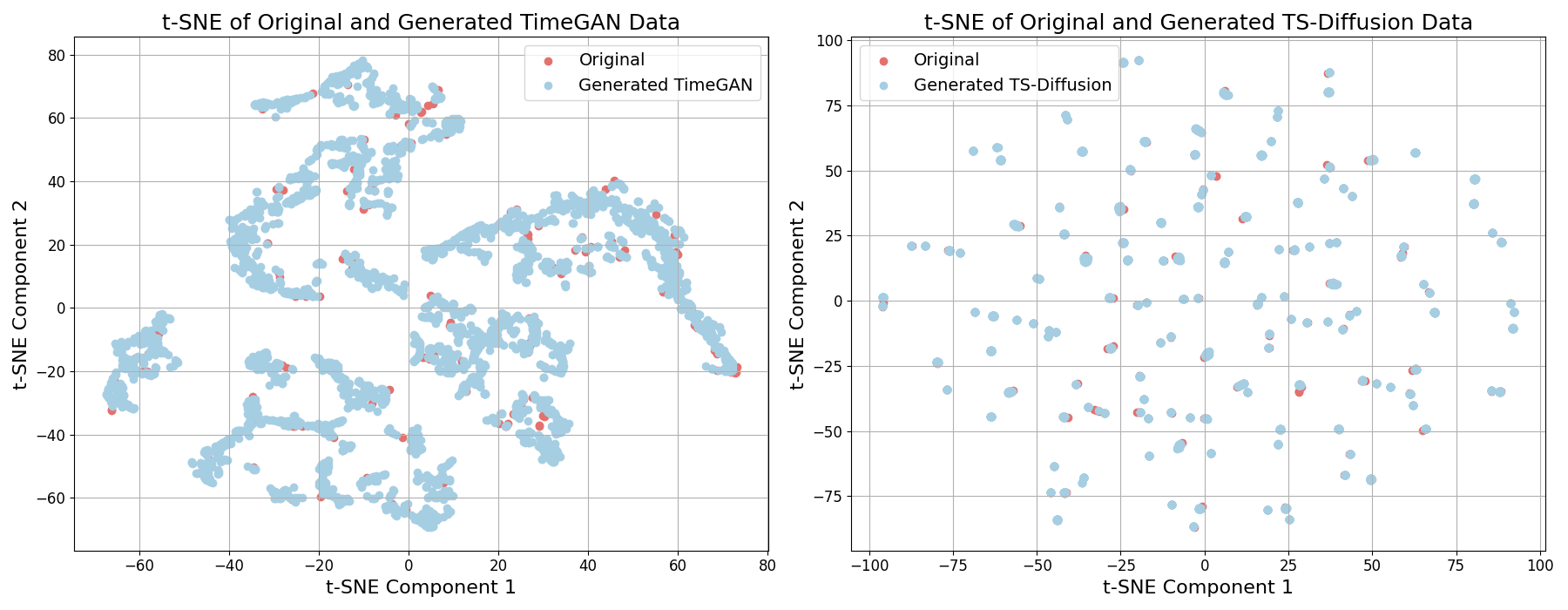}
  \caption{t-SNE of the TS-Diffusion generated data and the original data (right) and t-SNE of the TimeGAN generated data and the original data (left).}
  \label{fig5}
\end{figure}

Figures\ref{fig4} and \ref{fig5} show the PCA (Principal components analysis) plot versus t-SNE (t-Distributed Stochastic Neighbor Embedding) plot of the TS-Diffusion and TimeGAN generated data versus the original data.

From the PCA and t-SNE plots, it can be found that with an equal number of generated samples, the data generated by TS-Diffusion overlaps with the original data better. In contrast, the data generated by TimeGAN is relatively widely dispersed. Thus, we can infer that TS-Diffusion might learn the distribution of the input data better when generating the time series data, which is a key requirement in time series data generation.

\begin{figure}[htbp]
  \centering
  \includegraphics[width=1\linewidth]{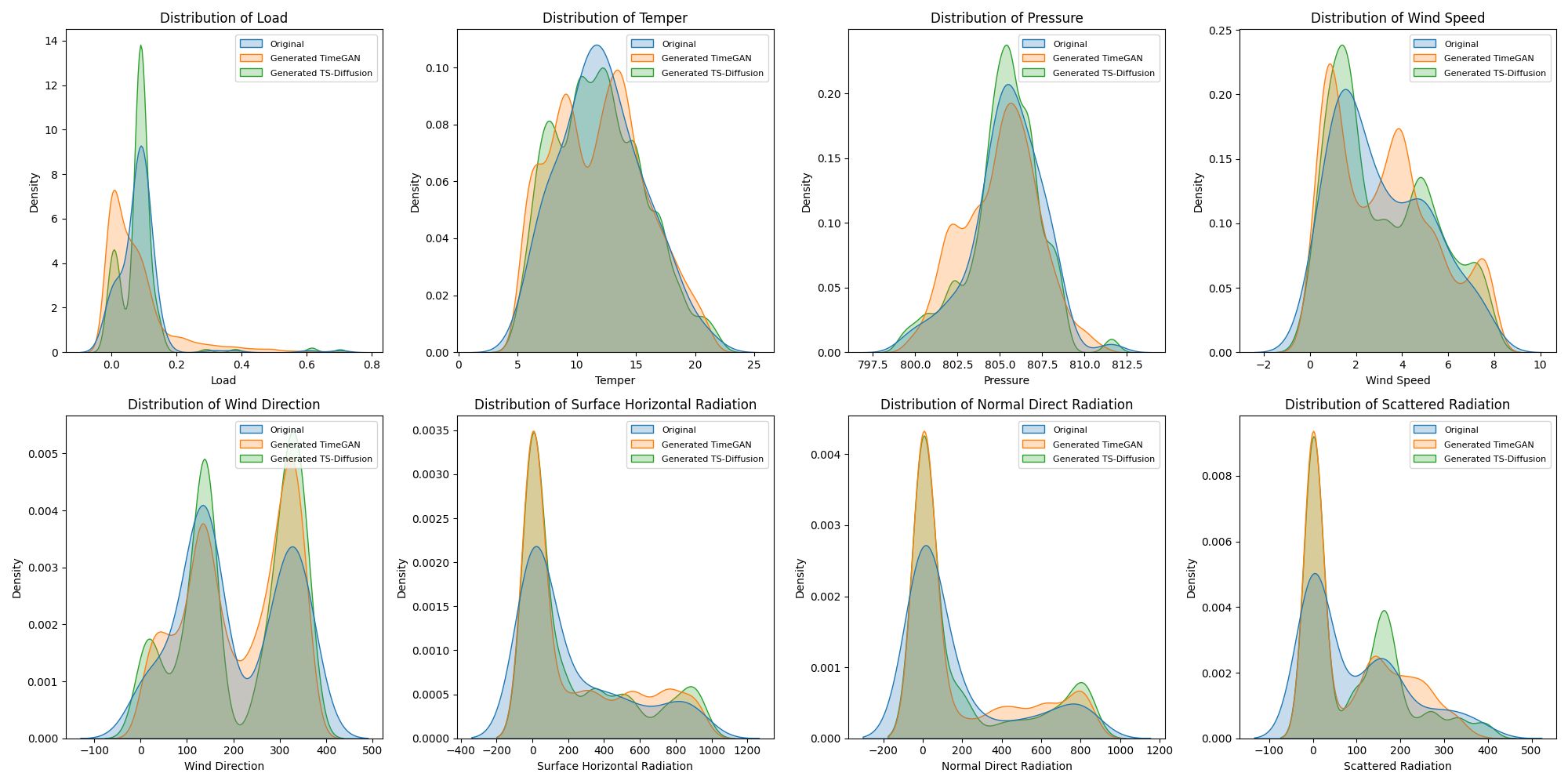}
  % \fbox{\rule[-.5cm]{0cm}{4cm} \rule[-.5cm]{4cm}{0cm}}
  \caption{KDE plot of TS-Diffusion augmented dataset, TimeGAN augmented dataset and original data.}
  \label{fig6}
\end{figure}

On the other hand, we also want to check if such a high overlap of the TS-Diffusion generated data with the original data on the above 2D plots means that the TS-Diffusion does not have the potential to \textit{dissimilate} the original data (Because nobody wants to always generate the same data as the original dataset!). So, we further inspect the distribution (i.e., probability density) of each generated feature column by the KDE (kernel density estimation) plots shown in Figure\ref{fig6}.

Finally, it can be observed that the distribution of the data generated by TS-Diffusion \textit{does not completely overlap} with the distribution profile of the original data (but still captures the basic position and shape of the original data). This further explains the previously observed superiority (Table~\ref{table1} vs. Table~\ref{table3}) of the TS-Diffusion augmented dataset over the TimeGAN augmented dataset. Briefly speaking, the TS-Diffusion model is \textit{not} \textit{simply} “copying” the target dataset when it tries to generate its own data.

In addition, we calculate the mean and standard deviation of the seven input features and the load demand for the original dataset and the augmented datasets, respectively, to inspect the quality of the augmented datasets from a statistical perspective. The results are shown in Table \ref{table4}.

\begin{table}
  \caption{Mean and standard deviation of three datasets}
  \label{table4}
  \centering
  \resizebox{\textwidth}{!}{
  \begin{tabular}{llclclc}
    \toprule
    Feature         &\multicolumn{2}{c}{Original dataset}    &\multicolumn{2}{c}{TimeGAN augmented}  &\multicolumn{2}{c}{TS-Diffusion augmented}\\
                     \cmidrule(r){2-3}                        \cmidrule(r){4-5}                       \cmidrule(r){6-7}        
                    &Mean     &Standard deviation    &Mean    &Standard deviation            &Mean    &Standard deviation \\
    \midrule
    Load            &0.0846   &0.0796                &0.0785  &0.1047                        &0.0840  &0.0819\\
    Temperature     &12.1151  &3.5519                &11.9075&3.8843                         &11.8959 &3.7428\\
    Pressure        &805.4252 &2.1036                &805.0396&2.2329                        &805.2451&2.1093\\
    Wind Speed      &3.0823   &2.0580                &3.2160&2.2103                          &3.1738  &2.1988\\
    Wind Direction  &200.2355 &112.9504              &209.9846&108.3983                      &211.6479&116.6709\\
    Surface Horizontal Radiation&223.0501&299.1725   &240.8833&311.9197                      &233.1411&311.5514\\
    Normal Direct Radiation&174.3329&280.0541        &199.4151&290.4430                      &194.1014&296.6413\\
    Scattered Radiation&95.8374&114.8145             &91.6883&107.3998                       &90.9040 &109.0043\\
    \bottomrule
  \end{tabular}
  }
\end{table}

It can be seen more clearly from Table \ref{table4} that although the augmented dataset is much larger, it still holds similar statistics as the original data, demonstrating again that the augmented models can indeed capture the original data's inherent patterns.

\end{document}